\documentclass[12pt]{iopart}

\newcommand{\eqb}{\begin{equation}}
\newcommand{\eqe}{\end{equation}}
\newcommand{\dmb}{\begin{displaymath}}
\newcommand{\dme}{\end{displaymath}}

\newcommand{\eab}{\begin{eqnarray}}
\newcommand{\eae}{\end{eqnarray}}

\begin{document}

\title[]{SU(2) Yang-Mills thermodynamics and photon physics}

\author{R Hofmann}

\address{Institut f\"ur Theoretische Physik, 
Universit\"at Karlsruhe (TH),  
Kaiserstr. 12,\\ 
76131 Karlsruhe, Germany}
\ead{hofmann@particle.uni-karlsruhe.de}
\begin{abstract}

Based on quantitative predictions enabled by a 
nonperturbative approach to Yang-Mills thermodynamics it is 
explained why the physics of photon {\sl propagation} is not unlikely 
rooted in pure SU(2) gauge dynamics.     

\end{abstract}

\pacs{70S15;74A15}
\vspace{2pc}
\noindent{\it Keywords}: thermal ground state, screening, photon
polarization, Planck-scale axion\\ 
\submitto{\JPA}

\section{Introduction}

Physics is the endeavor to understand in mathematical terms 
the fundamental laws governing our Universe. Quite generally, genuine 
progress in learning depends on the sophistication and perseverance 
in posing relevant questions. In physics the content of a question -- a
prediction -- is mathematically deduced from a prejudice 
(principle, postulate), and the prediction is either verified or falsified 
by experiment. The more experimentally verified, 
independent predictions there emerge without any falsification 
the more truth and generality is attributed to the starting principle. 

The purpose of this talk is to discuss 
implications of the postulate that the physics of photon 
propagation, which conventionally is 
associated with a U(1) gauge symmetry, is actually 
SU(2) Yang-Mills dynamics. While this 
may seem questionable and contrived judging 
by a counting of the perturbative degrees of freedom and their 
universal interactions a thermodynamic approach to SU(2) Yang-Mills
theory clearly suggests otherwise \cite{Hofmann2005B,SHG2006-1,SHG2006-2}. 
Namely, in the deconfining phase the gauge symmetry SU(2) is broken 
dynamically down to the subgroup U(1) by a 
nontrivial thermal ground state. The latter emerges upon a spatial 
coarse-graining  over interacting calorons and 
anticalorons of topological charge modulus $|Q|=1$
\cite{Hofmann2005}. While the thermal ground state is responsible for the
emergence of the (temperature-dependent) mass for two out of the three
species of gluons, that is, the dynamical gauge-symmetry breaking
SU(2)$\to$U(1), it also provides for a scale of maximal resolution 
$|\phi|$. The latter uniquely is determined by
temperature and the Yang-Mills scale and enforces in the effective
theory a rapidly converging loop expansion of thermodynamic
quantities. As a consequence, the polarization tensor
for the massless mode \cite{SHG2006-1}, computed on the one-loop level
yields a numerically reliable result for the modification of the
dispersion law, and contact with observation and 
experiment can be made once the above postulate is 
agreed upon.   

This presentation is organized as follows. In Sec.\,\ref{decsu2} we give
a brief summary of deconfining and 
preconfining Yang-Mills thermodynamics. The 
peculiarities of thermalized photon propagation 
in light of the above postulate are discussed 
in Sec.\,\ref{post}. In Sec.\,\ref{ev} we argue for a certain amount of
experimental evidence, and in Sec.\,\ref{con} we 
provide for an outlook on future activity.                 

\section{Deconfining and preconfining SU(2) Yang-Mills
  thermodynamics\label{decsu2}}

SU(2) Yang-Mills theory takes place in three 
distinct phases. At high temperature $T$ (deconfining phase) one shows 
\cite{Hofmann2005,HerbstHofmann2004,Hofmann2007} that an inert (nonfluctuating), adjoint
scalar field $\phi$ emerges upon a spatial coarse-graining over
interacting calorons and anticalorons of topological charge modulus
$|Q|=1$. Performing this coarse-graining over a trivial-holonomy
caloron-anticaloron pair in singular gauge and resorting to a particular
global gauge choice, one has 
\eqb
\label{phiwind}
\phi=2\,\sqrt{\frac{\Lambda^3\beta}{2\pi}}\,t_1\,\exp(\pm\frac{4\pi
  i}{\beta}t_3\tau)\,,
\eqe
where $\Lambda$ is a purely nonperturbative constant of integration (the
Yang-Mills scale) \cite{GiacosaHofmann2006}, 
$0\le\tau\le\beta\equiv\frac{1}{T}$ is the euclidean 
time, and $t_a$ are SU(2) generators in the fundamental representation 
normalized as tr\,$t_at_b=\frac12\delta_{ab}$. The entire effective
action (including the coarse-grained sector of topologically
trivial field configurations) follows from perturbative 
renormalizability \cite{'tHooftVeltman} and gauge invariance, and the 
thermal ground state is given by Eq.\,(\ref{phiwind}) and the pure-gauge 
configuration $a_\mu^{\tiny\mbox{bg}}=\mp\delta_{\mu
  4}\frac{2\pi}{e\beta}\,t_3$. Here $e$ is the effective 
gauge coupling whose evolution with temperature is 
determined by the Legendre transformations in the 
effective theory. This evolution possesses an attractor: Evolving
downward in temperature, $e$ rapidly approaches the 
plateau $e=\sqrt{8}\pi$ for $\lambda\equiv\frac{2\pi
  T}{\Lambda}\gg\lambda_c=13.87$ and runs into a pole of the form 
$e\propto-\log(\lambda-\lambda_c)$. Here $T_c$ is the temperature where
totally screened magnetic monopoles start to condense. 
The ground-state pressure $P^{\tiny\mbox{gs}}$ is
negative: $P^{\tiny\mbox{gs}}=-4\pi\Lambda^3 T$. By an admissible change 
of gauge, such that $\phi\equiv
2\,\sqrt{\frac{\Lambda^3\beta}{2\pi}}\,t_3$ and
$a_\mu^{\tiny\mbox{bg}}=0$, the adjoint Higgs mechanism manifestly 
generates quasiparticle masses for the topologically trivial 
gauge fields $a^{1,2}_\mu$ while the field $a^3_\mu$ 
remains massless. Radiative corrections to thermodynamic quantities are
small even though the plateau value of $e$ is not small. They are 
computed in a loop expansion in the effective theory 
\cite{Hofmann2006}. This expansion converges rapidly because of infrared 
stability enabled by quasiparticle masses on tree level 
and because of kinematic constraints due 
to the existence of the maximal resolution $|\phi|$. In particular, it
is sufficient for practical purposes to compute the polarization tensor
of the massless mode to one-loop accuracy only \cite{SHG2006-1}. Depending
on their frequency, there is screening or antiscreening  
in a thermal gas of massless particles due to scattering involving the 
massive modes. Because of the dynamical gauge symmetry breakdown
SU(2)$\longrightarrow$ U(1) it is tempting to attribute the existence
and propagation of the photon to this Yang-Mills theory. In the
preconfining phase, that is, for $T$
slightly below $T_c$ magnetic\footnote{Magnetic w.r.t. the defining
  SU(2) Yang-Mills theory, electric w.r.t. photons.} monopoles start to condense. 
In spatial regions, where a stable condensate prevails, the unbroken 
U(1) symmetry of the deconfining phase is dynamically broken. For the
photon this would mean that an additional polarization emerges if 
temperature falls below $T_{\tiny\mbox{CMB}}$.   

Because no screening or antiscreening is observed for
long-wavelengths photons emitted by astrophysical sources and  
propagating towards Earth above the present ground 
state of the cosmic microwave 
background (CMB) and because this is the situation predicted at $T_c$ 
by an SU(2) Yang-Mills theory\footnote{At $T_c$ massive
  quasiparticles decouple thermodynamically and thus do 
not contribute to screening or antiscreening of the massless mode
\cite{SHG2006-1,SHG2006-2,HHR2004}.} we are led to identify $T_c$ with the
present value of $T_{\tiny\mbox{CMB}}\sim 2.73\,$K. This, in turn, fixes
the Yang-Mills scale as $\Lambda_{\tiny\mbox{CMB}}=2.35\times
10^{-4}\,$eV.    

\section{The postulate
  SU(2)$_{\tiny\mbox{CMB}}\stackrel{\tiny\mbox{today}}=$U(1)$_{\tiny\mbox{photon}}$\label{post}}

\begin{figure}
\begin{center}
\leavevmode
\leavevmode
\vspace{4.9cm}
\includegraphics{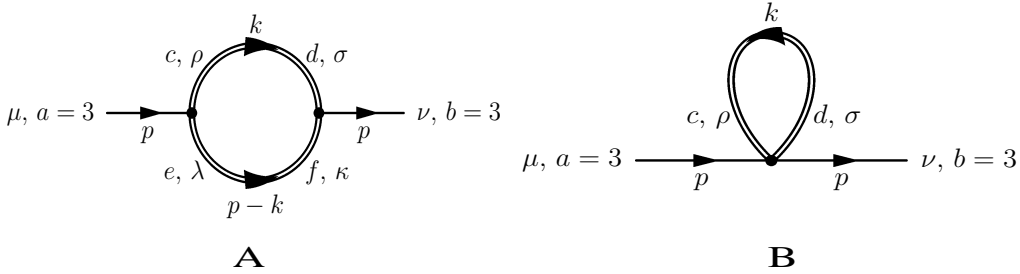}
\end{center}
\caption{\protect{\label{Fig-1}} The diagrams for the TLM mode polarization tensor.}      
\end{figure}

\begin{figure}
\begin{center}
\leavevmode
\leavevmode
\vspace{6.5cm}
\includegraphics{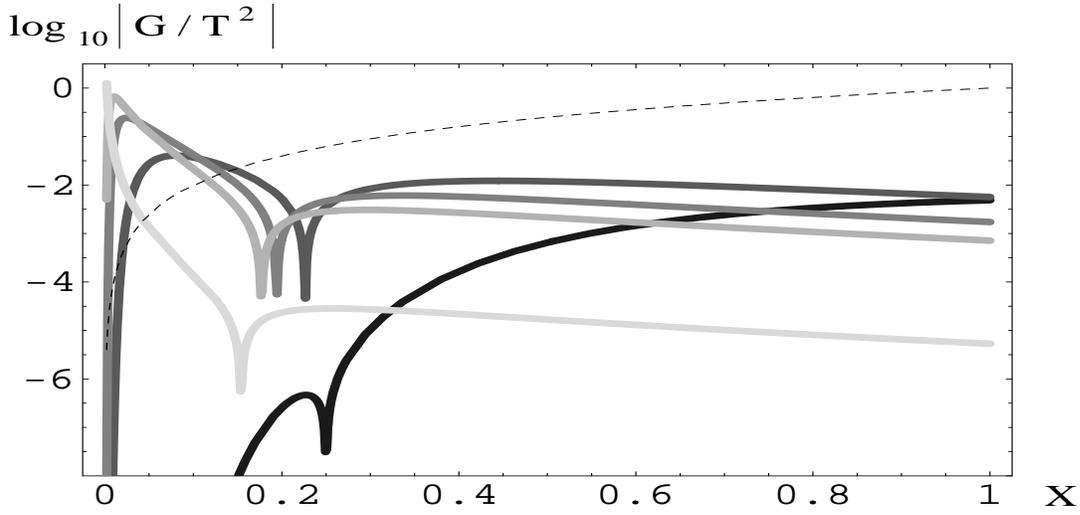}
\end{center}
\caption{\protect{\label{Fig-2}} A plot of $\log \frac{|G|}{T^2}$ as a
  function of $X$ for $\lambda=1.12\,\lambda_c$ (black),
  $\lambda=2\,\lambda_c$ (dark grey), $\lambda=3\,\lambda_c$
  (grey),$\lambda=4\,\lambda_c$ (light grey), $\lambda=20\,\lambda_c$
  (very light grey). The dashed curve depicts the function
  $f(X)=2\log_{10}\,X$. }      
\end{figure}

Subjecting photon propagation to an SU(2) 
gauge principle we refer to the two massive modes as $V^\pm$ and, as usual, 
to the massless excitation as $\gamma$. Screening or antiscreening of
thermalized $\gamma$-radiation is a small effect for thermodynamic
quantities such as the pressure \cite{SHG2006-1} 
which peaks at about twice $T_c$. Depending on its frequency $\omega$
and spatial momentum $\vec{p}$, 
the modification of the U(1) dispersion law is as 
\eqb
\label{moddisplaw}
\omega^2=\vec{p}^2 \longrightarrow
\omega^2=\vec{p}^2+G(\omega,|\vec{p}|,T,\Lambda_{\tiny\mbox{CMB}})\,.
\eqe
The function $G$ enters the polarization tensor $\Pi_{\mu\nu}$. For
$\omega=|\vec{p}|$ the function $G$ is real, corresponds to $\Pi_{11}=\Pi_{22}$ if $\vec{p}$ points
into the 3-direction, and is computed according\footnote{If the condition
  $\omega=|\vec{p}$ is sizably modified then also Feynman diagram A in
  in Fig.\,\ref{Fig-1} contributes, and $G$ acquires an imaginary part.} to the Feynman diagram B
in Fig.\,\ref{Fig-1}.
In Fig.\,\ref{Fig-2} the dependence of $G$ on dimensionless 
momentum $X\equiv\frac{|\vec{p}|}{T}$ is depicted for various
temperatures\footnote{Setting $\omega=|\vec{p}|$ in $G$ makes it a
  function of $\omega$ only. This approximation turns out to be 
selfconsistent \cite{SHG2006-2} for almost all values of $\omega$.}. 
To the left (right) of the cusps $G$ is positive (negative) corresponding
to screening (antiscreening). Points lying above the dashed curve are
associated with strongly screened modes (screening mass larger than
modulus of momentum). 

\begin{figure}
\begin{center}
\leavevmode
\leavevmode
\vspace{4.9cm}
\includegraphics{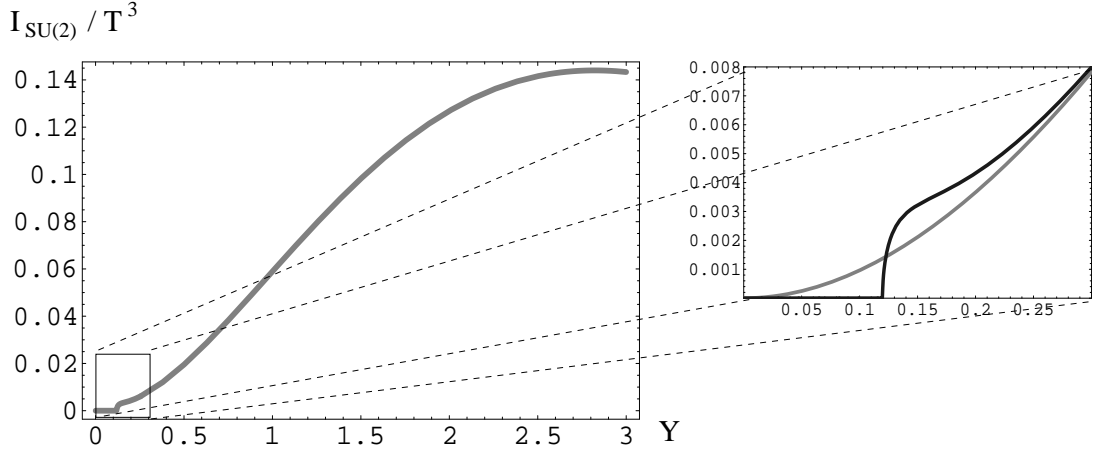}
\caption{\protect{\label{Fig-3} Dimensionless black-body spectral power 
$\frac{I_{\tiny\mbox{SU(2)}}}{T^3}$ as a function of the dimensionless frequency 
$Y\equiv\frac{\omega}{T}$. The black curve in the magnified region depicts the modification of 
the spectrum as compared to $\frac{I_{\tiny\mbox{U(1)}}}{T^3}$ (grey curve) 
for $T=10\,$K.}} 
\end{center}     
\end{figure}

What are the implication of the modification in 
Eq.\,(\ref{moddisplaw}) for the black-body spectrum? The total energy density
$\rho$ of a thermal gas of $\gamma$ is defined as
\eqb
\label{rhodef}
\rho\equiv
2\int\frac{d^3p}{(2\pi)^3}\,\omega\,n_B\left(\frac{\omega}{T}\right)\,,
\eqe
where $n_B(x)\equiv\frac{1}{\exp[x]-1}$ denotes the Bose
distribution. Expressing the momentum-space measure $d^3p$ 
in terms of a frequency measure under consideration of the 
modified dispersion law in Eq.\,(\ref{moddisplaw}), one has
\eqb
\label{meastrafo}
\int d^3p=4\pi\int d|\vec{p}|\,|\vec{p}|^2=4\pi\int
d\omega\,\sqrt{\omega^2-G(\omega)}\,\left(\omega-\frac12\frac{dG(\omega)}{d\omega}\right)\,,
\eqe
where the additional dependence of $G$ on $T$ is suppressed. In the
strong-screening regime the quantity $|\vec{p}|$ would be imaginary, and
thus the integration over $\omega$ is restricted to a domain where
$\omega^2\ge G$. Thus we can write the spectral intensity $I_{\tiny\mbox{SU(2)}}(\omega)$ of the 
SU(2)-modified black body in terms of the spectral intensity
$I_{\tiny\mbox{U(1)}}(\omega)$ 
of the U(1) black body as  
\eqb
\label{PmBB}
I_{\tiny\mbox{U(1)}}(\omega)\to I_{\tiny\mbox{SU(2)}}(\omega)=I_{\tiny\mbox{U(1)}}(\omega)\times
\frac{\left(\omega-\frac{1}{2}\frac{d}{d\omega}G\right)\sqrt{\omega^2-G}}{\omega^2}\,
\theta(\omega-\omega^*)\,,
\eqe
where $\omega^*$ is the root\footnote{There are actually two roots,
  compare with Fig.\,\ref{Fig-2}. For many practical concerns 
the lower lying root can safely be set equal to zero.} 
of $\omega^2=G$, $\theta(x)$ is the Heaviside step 
function, and 
\eqb
\label{PiBB}
I_{\tiny\mbox{U(1)}}(\omega)=\frac{1}{\pi^2}\,\frac{\omega^3}{\exp[\frac{\omega}{T}]-1}\,.
\eqe
Fig.\,\ref{Fig-3} depicts the modified black-body spectrum according to
Eq.\,(\ref{PmBB}) at $T=10\,$K. 
For $T<T_c=T_{\tiny\mbox{CMB}}$ $\gamma$ starts to acquire a Meissner
mass, and the average number of photon polarizations rapidly increases
from two towards three. This, in turn, implies a rapid increase of the
energy density of the photon gas as compared to its value in the U(1)
theory.  

\section{Evidence in nature?\label{ev}}

\begin{figure}[tbp]
\begin{center}
\leavevmode
\leavevmode
\vspace{10.5cm} \includegraphics{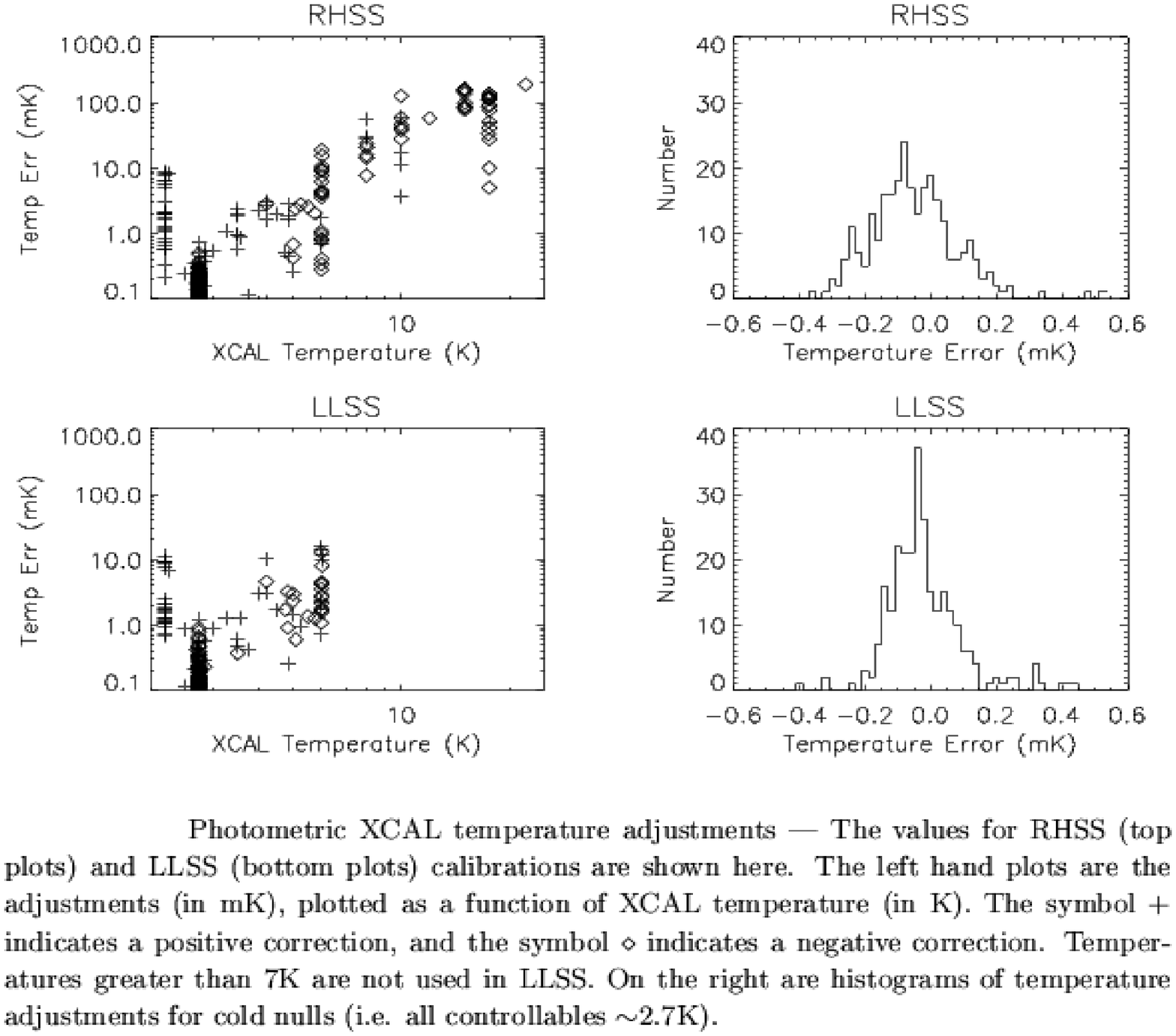}
\caption{Temperature offsets as measured in the FIRAS orbit
  calibration. Notice the peak at $T=2.2\,$K. 
Figure taken from \protect{\cite{FIRASdoc}}.}
\end{center}
\label{Fig-4}
\end{figure}

In \cite{SHGS} an analysis of the predictions of
SU(2)$_{\tiny\mbox{CMB}}$ for temperatures offsets\footnote{Defined by $\delta T\equiv T_{\tiny\mbox{rad}}-T_{\tiny\mbox{XCAL}}$,
  where $T_{\tiny\mbox{rad}}$ is 
extracted by fitting $I_{\tiny\mbox{U(1)}}$ to the 
intensity of the radiation and $T_{\tiny\mbox{XCAL}}$ is the (wall) 
temperature of the calibrator.} $\delta T$ was performed along the lines of the 
COBE Firas situation. Their data of the spectral shape of the 
black-body intensity for temperatures in the vicinity of
$T_{\tiny\mbox{CMB}}=2.73\,$K was taken during the calibration 
stage of the instrument \cite{FIRASdoc}. 
A comparison of their temperature 
offsets $\delta T$ with the predictions of 
SU(2)$_{\tiny\mbox{CMB}}$ reveals 
that the predicted anomaly is smaller than the 
experimental error in the FIRAS calibration. What is interesting,
however, is the sudden increase of $\delta T$ for
$T_{\tiny\mbox{XCAL}}<T_{\tiny\mbox{CMB}}$, see Fig.\,\ref{Fig-4}, which
we attribute to an increase of the average number of photon 
polarizations at the onset of the preconfining phase, for a discussion
see \cite{SHGS}.
Next there is large-angle suppression 
in the CMB TT power spectrum and the statistical correlation of the 
low multipoles \cite{Schwarz2007}. Based on the black-body anomaly
predicted by SU(2)$_{\tiny\mbox{CMB}}$ a model for the generation of
large-angle temperature fluctuations in the CMB was proposed in
\cite{SH2007} which has the potential to explain these effect in 
terms of a large dynamical contribution to the CMB dipole, see also 
\cite{SHGS}. Third, large, old, cold, and dilute clouds of 
atomic hydrogen were discovered in between spiral arms of the outer 
Milky Way, see \cite{BruntKnee2001}. The puzzling fact about these clouds is
their inferred age of about 50 million years. This is much older than 
model calculations for the duration for the formation of 
sizable fractions of H$_2$ molecules suggest, for one of the newer
investigations see \cite{Goldsmith}. In \cite{SHG2006-2} it was pointed
out that the interatomic distance of about 1\,cm 
between the hydrogen atoms is roughly equal to the 
wavelength of screened photons at the relevant (brightness) temperatures 
of 5\,K to 10\,K, and that the 21\,cm--line, which thermalizes the cloud 
system, propagates. By computing the two-point correlator of the photon 
energy density, this observation is confirmed \cite{KHG2007}. That is, 
photons, needed to mediate interactions between the hydrogen atoms, are
screened due to the nonabelian effects of SU(2)$_{\tiny\mbox{CMB}}$, and
the cloud changes its composition on a much slower rate than 
conventionally expected. Fourth, a scenario was discussed in
\cite{GH2005} where the nontrivial thermal ground state of
SU(2)$_{\tiny\mbox{CMB}}$, by virtue of dynamical chiral symmetry
breaking \cite{BC} and the chiral anomaly \cite{ABJ} invoked at the Planck scale, gives 
rise to an ultralight axion field. If CP violating signatures, such as a
nonvanishing EB cross correlation at large angles, 
will be discovered in future CMB satellite missions, then this Planck-scale
axion field would yield a theoretically and observationally backed up 
explanation of the present cosmological concordance model. 
That is, the physics of visibility (propagating photons) would be
unified with the physics of darkness (dark matter and dark energy) in
terms of an SU(2) gauge principle. 

Finally, let us discuss an apparent puzzle: Even at room 
temperature a sizable fraction of the radiowave spectrum is screened 
according to the modified dispersion law in Eq.\,(\ref{moddisplaw}). 
But we do not observe this screening in our daily broadcasts. 
So why is this? The answer is that the intensity in a beam of 
radiowaves of a given frequency, as transmitted by a commonly used 
antenna, is by orders of magnitude 
larger than its corresponding black-body intensity. Thus those
radiowaves are a priori
not part of the thermal black-body 
spectrum at, say, room temperature. The question then arises
how long it takes for radiowaves to decrease their energy by radiating 
off $V^\pm$ particles to eventually be part of the thermal 
spectrum. The rate for this process is determined by the imaginary part
of a two-loop diagram (involving two four-vertices) for the polarization 
tensor. Since the real part of a two-loop diagram generally is suppressed by a factor of
$\sim 10^{-3}$ \cite{SHG2006-1} as compared to the one-loop result and 
since there is an even greater suppression for the imaginary part 
we expect no adulteration of radiowave propagate over terrestial
distances as compared to the U(1) theory. 
Recall, that there is no screening effect or energy loss whatsoever for
photon propagation above the present CMB ground state (radiowave
propagation in space) due to the thermal decoupling of 
$V^\pm$ at $T_c=T_{\tiny\mbox{CMB}}=2.73\,$K.                      
      
\section{Conclusions and outlook\label{con}}

In this talk we have given a brief account of why deconfining SU(2)
Yang-Mills thermodynamics may be the theory underlying 
photon propagation. We have mentioned evidence in favor of 
this postulate. A conclusive judgement will, however, 
be provided by a direct terrestial measurement of the 
spectral intensity of a low-temperature (say, $T=5\,$K to $T=10\,$K)
black body at low frequencies. If the spectral gap, as predicted by 
SU(2)$_{\tiny\mbox{CMB}}$, indeed is seen in a precision black-body 
experiment then this would imply far-reaching consequences 
for our understanding of electroweak symmetry breaking, for a discussion
see \cite{SHG2006-2}.  

Some of our future activity will be focussing on predictions of the average number
of photon polarizations in the supercooled, finite-volume situation.

\end{document}